\documentclass[reprint,fleqn,superscriptaddress,twocolumn,showpacs,amsmath,amssymb,prb,aps,longbibliography]{revtex4-1}
\pdfoutput=1
\usepackage[all]{xy}
\usepackage{gensymb}
\usepackage{graphicx,psfrag,times,epsfig,color}
\usepackage{verbatim,natbib}

\usepackage{makeidx}
\usepackage{amsmath}
\usepackage{bm}
\usepackage{amsfonts}
\usepackage{amssymb}
\usepackage{hyperref}

\begin{document}

\title{Spatial structure of correlations around a quantum impurity at the edge of a two-dimensional topological insulator}
\author{Andrew Allerdt}
\affiliation{Department of Physics, Northeastern University, Boston, Massachusetts 02115, USA}
\author{A.~E.~Feiguin}
\affiliation{Department of Physics, Northeastern University, Boston, Massachusetts 02115, USA}
\author{G.~B.~Martins}
\email[Corresponding author: ]{martins@oakland.edu}
\affiliation{Department of Physics, Oakland University, Rochester, MI 48309, USA}
\affiliation{Instituto de F\'{\i}sica, Universidade Federal Fluminense, 24210-346 Niter\'oi, RJ, Brazil}

\begin{abstract}
We calculate exact zero-temperature real space properties of a substitutional
magnetic impurity coupled to the edge of a zigzag silicene-like nanoribbon. 
Using a Lanczos transformation [Phys. Rev. B {\bf 91}, 085101 (2015)] and the density matrix 
renormalization group method, we obtain a realistic description of 
stanene and germanene that includes the bulk and the 
edges as boundary one-dimensional helical metallic states.
Our results for substitutional impurities indicate that the development of a Kondo state and the structure of the 
spin correlations between the impurity and the electron spins in the metallic edge state depend considerably on 
the location of the 
impurity. More specifically, our real space resolution allows us to conclude that
there is a sharp distinction between the impurity being located at a crest or a trough site at the zigzag edge.
We also observe, as expected, that the spin correlations are 
anisotropic due to an emerging Dzyaloshinskii-Moriya interaction with the conduction electrons, 
and that the edges scatter from the impurity and ``snake'' or circle around it. 
Our estimates for the Kondo temperature indicate that there is a very 
weak enhancement due to the presence of spin-orbit coupling.
\end{abstract}

\maketitle

\section{Introduction} 

A revolution is underway in the study of two-dimensional (2d) materials. 
Since the mechanical exfoliation of graphene from graphite was achieved 
in 2004 \cite{Novoselov2004} [see Ref.~\onlinecite{CastroNeto2009} for a detailed review on graphene properties], 
many resources have been invested in the synthesis of other monolayer 
systems. These efforts have been rewarded through the discovery 
of many new 2d compounds that are either stable in free-standing form 
or grown in a substrate platform [for a 
comprehensive review, see Ref.~\onlinecite{Butler2013}]. Among these new 2d materials, 
a few raise additional possibilities linked to the presence of non-trivial topological phases
such as silicene, germanene, and stanene. As a 
result of the spin-orbit interaction (SOI), they present bulk-gapped phases 
with  metallic helical (spin-momentum locked) 
edge states [for a review, see Ref.~\onlinecite{Ezawa2015}]. The locking of 
spin and momentum leads to suppression of elastic backscattering of the helical 
electrons in the absence of spin-flip processes. This, in turn, leads to the possibility 
of dissipationless spin polarized currents (see Refs.~\onlinecite{Konig2008,Hasan2010b,Qi2011b} 
for comprehensive reviews).

Besides the interest in the basic physics associated to topological phases in 
condensed matter materials (either 2d or 3d), the recent explosion of work in this field 
is also due to possible applications of topological insulators (TIs) 
in spintronics\cite{Fan2014, Hoffmann2015, Fong2016}. For instance, strong 
spin-transfer-torque effects have recently been observed at room temperature in a 3d TI \cite{Mellnik2014}.

The search for new paradigms in electronics has spurred interest in 
other quantum phenomena at the nanoscale: 
electron correlations offer the promise of new functionality in nanoelectronics 
related to the possible manipulation of many-body ground states in suitably 
produced nanostructures such as quantum dots and nanoribbons \cite{Andergassen2010a,Brus2014,Laird2015}. 
An important example that has attracted a great deal of attention is the Kondo state, realized by coupling 
a magnetic impurity to conduction electrons 
 [see, for example, section VII of Ref.~\onlinecite{Laird2015} 
for a recent review of the Kondo effect in carbon nanotubes]. 
In addition, Kondo physics offers the possibility 
of probing the spin-texture surrounding the magnetic moment,
and to gain valuable information about the effect 
of magnetic interactions over the metallic surface state \cite{note1}. 

A brief review of the Kondo effect in TIs (with a focus on quantum critical behavior) can be found in Ref.~\onlinecite{Vojta2015}. 
Initial experimental work concentrated on bulk-doped 3d TIs
 \cite{Chen2010, Hor2010,Chang2013}, but since many TI-based devices require the surface of the TI to be 
in contact with ferromagnets, 
experiments were also done involving {\it surface} deposition of magnetic impurities in 3d TIs 
\cite{Wray2011,Yu2014,Wang2015a,Scholz2012,Honolka2012}. 

In a theoretical study of the Kondo effect at the surface of a 3d 
TI, {\v{Z}}itko \cite{Zitko2010} found that 
the Hamiltonian of a quantum magnetic impurity coupled to metallic topological surface states 
maps into a conventional pseudogap single-channel Anderson impurity model with SU(2) symmetry. 
It was also pointed out that despite the relatively trivial nature of the low-energy Hamiltonian, 
the screening Kondo cloud should possibly display a rather complex structure, which would be 
reflected in non-trivial spatial dependencies of the spin correlations between the magnetic impurity 
and the topological surface states involved in the Kondo-singlet formation. 

Motivated by Fourier-transform scanning tunneling spectroscopy 
measurements done in Iron-doped ${\rm Bi_2Te_3}$ \cite{Okada2011}, 
which analyzed the energy-dependent spatial variations of 
the local density of surface states in terms of quasiparticle interference (QPI), 
Mitchell {\it et al.} \cite{Mitchell2013a} conducted numerical simulations of a single 
Kondo impurity on the surface of a 3d TI to  
identify the signatures in QPI of the Kondo
interaction between the helical metal and the impurity. 
The QPI simulation results were found to be 
markedly different from those obtained for 
nonmagnetic or static magnetic impurities.

%%%%%%%%%%%%%%%%%%%%%% 
The influence of the SOI on the Kondo effect in 2d TIs has been addressed 
in several numerical works \cite{Zitko2011, Zarea2012, Mastrogiuseppe2014, Wong2016, Vernek2016}, 
particularly focusing on the behavior of the Kondo temperature. In overall agreement with 
Ref.~\onlinecite{Zitko2010}, Isaev {\it et al.} \cite{Isaev2015} demonstrated that strong SOI 
leads to an unconventional Kondo 
effect (despite being of the SU(2) kind) with an impurity spin screened by purely
orbital motion of surface electrons. At low energies, the impurity 
spin forms a singlet state with the
total electron angular momentum, and the system exhibits an
emergent SU(2) symmetry, which is responsible for the Kondo resonance.

Quantum Monte Carlo (QMC) simulations \cite{Hu2013} applied to a single impurity Anderson model 
in a zigzag graphene edge analyzed the influence of spin-momentum 
locking over the Kondo state and found indications of 
a broken spin rotation symmetry in spin-spin correlation functions. A limitation 
of this effort, stemming from constraints associated to the QMC method, is the high value of 
the minimum temperature achieved ($\approx 1000~{\rm K}$ for graphene). 

In this work, 
we present results for the spin correlations around a substitutional Anderson impurity at 
the edge of a silicene-like topological insulator (more specifically for germanene and stanene, 
see Table \ref{table1}), with full spatial resolution of the lattice. To the best of our knowledge, 
a detailed study of these correlations in the Kondo ground state has not been attempted so far. 

The structure of this article is as follows: In Sec.~\ref{model} we present the model and 
a very brief description of the numerical method used (for details  
the reader is referred to previous papers by the authors \cite{Busser2013a,Allerdt2015,Allerdt2016}). 
In Sec.~\ref{ribbon-band} we present the band structure for a zigzag nanoribbon (ZNRB) in the TI phase, 
focusing on the edge states. Section \ref{ldos} presents local density of states 
(LDOS) results for sites at the edge of the ZNRB, showing how it changes with SOI 
compared to the LDOS of bulk sites.  Section \ref{sscorr} shows results for spin correlations between the localized magnetic 
moment and the conduction spins.
We analyze the cases of a substitutional impurity sitting at either crest or trough sites (as defined is 
Sec.~\ref{model}), and study the effects of spin orbit coupling. An analysis of the influence of the SOI 
on the Kondo temperature is performed in Sec.~\ref{stk}. The paper closes with a summary and conclusions.

\section{Model}\label{model}
The independent electron Hamiltonian $H_{\rm band}$   
describing the 2-dimensional topological insulator corresponds to a tight-binding band structure that 
is appropriate for silicene, germanene, and stanene \cite{Ezawa2015}: 
\begin{flalign}
H_{\mathrm{band}} &=-t\sum_{\left\langle i,j\right\rangle \sigma}c_{i\sigma}^{\dagger }c_{j\sigma}
+i\frac{\lambda_{\text{SO}}}{3\sqrt{3}}\sum_{\left\langle \!\left\langle
	i,j\right\rangle \!\right\rangle \sigma}\sigma\nu _{ij}c_{i\sigma}^{\dagger }c_{j\sigma}, &
\label{HamilSilic}
\end{flalign}
where $c_{i\sigma}^{\dagger }$ creates an electron in site $i$ with spin $\sigma$ 
(note that the letter $\sigma$ stands for $ \sigma=\uparrow \downarrow $ 
when used as a subindex, and for $\sigma=\pm $ 
when used within equations). In addition, $\left\langle i,j \right\rangle $ runs over nearest-neighbor sites and 
$\left\langle \!\left\langle i,j\right\rangle \!\right\rangle $ runs 
over next-nearest-neighbor sites. The first term describes nearest-neighbor hoppings 
with transfer integral $t$. The second term is the effective spin-orbit interaction 
with coupling $\lambda_\text{SO}$, where $\nu_{ij}=+1$ if the next-nearest-neighbor hopping is anticlockwise and 
$\nu_{ij}=-1$ if it is clockwise (in relation to the positive $z$-axis). 
The parameter values for silicene, germanene, and stanene are given in Table \ref{table1}, where the corresponding 
values for graphene are given for comparison (note that as graphene does not buckle, 
its $\lambda_\text{SO}$ value effectively vanishes, thus it is gapless and has no measurable 
topological properties, contrary to silicene, germanene, and stanene). In Table \ref{table1} we also 
list the Rashba spin-orbit interaction for each material. For the sake of simplicity 
(and given its small value) we omitted it from our calculations. The unit of energy for all results shown here is $t$.  

In accordance with Table \ref{table1}, the ratios $\lambda_\text{SO}/t$ for silicene, germanene, and 
stanene are $0.0024$, $0.033$, and $0.077$. We will use in our calculations (unless stated otherwise) 
a value three times larger than the one for germanene, i.e., $\lambda_\text{SO} = 0.1$. 
Therefore, our results should describe the Kondo effect in germanene
and stanene. Due to its much smaller SOI, silicene is expected to have a 
behavior (not shown) very similar to that of graphene. 

\begin{table}[tbp]
\begin{center}
\begin{tabular}{|l|l|l|l|l|l|l|l|}
\hline
& $t$(eV) & $v_F$ & $a$(\r{A}$)$ & $\lambda_{\text{SO}}$ & $\lambda_{\text{R}}$ & $\theta $ \\ \hline
Graphene & 2.8 & 9.8 & 2.46 & 10$^{-3}$ & 0 & 90 \\ \hline
Silicene & 1.6 & 5.5 & 3.86 & $3.9$ & 0.7 & 101.7 \\ \hline
Germanene & 1.3 & 4.6 & 4.02 & $43$ & 10.7 & 106.5 \\ \hline
Stanene & 1.3 & 4.9 & 4.70 & $100$ & 9.5 & 107.1 \\ \hline
\end{tabular}
\end{center}
\caption{The parameters characterizing graphene, silicene, germanene, and stanene. The Fermi velocity 
$v_{\text{F}}$ is in units of 10$^{5}$m/s, $\protect\lambda_{\text{SO}}$ 
and $\protect\lambda_{\text{R}}$ (Rashba SOI) in meV. $\protect\theta $ is the bond angle. 
Adapted from Ref.~\protect\onlinecite{Ezawa2015} (see also Ref.~\protect\onlinecite{Liu2011a}).}
\label{table1}
\end{table}

The total Hamiltonian $H_{\mathrm{T}} = H_{\mathrm{band}} + H_{\mathrm{imp}} + H_{\mathrm{hyb}}$ includes, besides $H_{\mathrm{band}}$, also 
the impurity and its hybridization with the lattice: 
\begin{eqnarray}
H_{\mathrm{imp}} &=& \epsilon_0 (n_{\mathrm{imp},\uparrow}+n_{\mathrm{imp},\downarrow}) + U n_{\mathrm{imp},\uparrow} n_{\mathrm{imp},\downarrow} \\
	H_{\mathrm{hyb}} &=& V \sum_\sigma (c_{\mathrm{imp},\sigma}^{\dagger} c_{k\sigma} + \mathrm{h.c.}),
\end{eqnarray}
where $n_{\mathrm{imp},\sigma}=c_{\mathrm{imp},\sigma}^{\dagger}c_{\mathrm{imp},\sigma}$ and $c_{\mathrm{imp},\sigma}^{\dagger}$ ($c_{\mathrm{imp},\sigma}$) 
creates (annihilates) an electron at the impurity, which is described by a single impurity Anderson model (SIAM) 
with Coulomb interaction $U$ and orbital energy $\epsilon_0$. 

As for the hybridization between the impurity and the lattice, 
$H_{\mathrm{hyb}}$, we will focus on a {\it substitutional} impurity, which replaces an atom from either sublattice and therefore 
has overlap integrals with more than one lattice site (from the opposite sublattice). 
Therefore, $c_{k\sigma}$ stands for a symmetric linear combination of two (or three, 
depending on which edge site we are considering) nearest-neighbors to the lattice site occupied by the impurity. 
In the case of a zigzag nanoribbon (studied in this work), with an edge geometry schematically represented as 
$\cdots /\backslash/\backslash \cdots$, which we denote as a sequence of sites $\cdots ABABA \cdots$, 
the choice is between $A$ sites (with coordination three) and $B$ sites (with coordination two). 
Our calculations show a wide variation in the results (mainly for the spin correlations) depending 
on what edge site the impurity replaces. 
In the following, we will refer to $A$ and $B$ sites in the zigzag profile as `trough' 
and `crest' sites, respectively. Our results clearly show that the metallic topologically nontrivial states 
that screen the magnetic impurity reside in the crest sites, while the trough sites and their immediate neighborhood, 
as our spin correlation results have shown, behave as small metallic ``puddles'', which leak from the metallic edges, 
and are surrounded by the bulk. 

In order to perform unbiased numerical simulations of the model just described 
to calculate spin-spin correlations, we use the so-called block Lanczos method  recently introduced by the authors \cite{Allerdt2015,Allerdt2016} 
(see also Ref.~\onlinecite{Yunoki.Block} for 
an independent development of the same ideas). This approach enables one 
to study quantum impurity problems with a real space representation of the lattice, and in 
arbitrary dimensions, using the density matrix renormalization group method (DMRG)\cite{White1992,White1993}.
By generalizing the ideas introduced in Ref.~\onlinecite{Busser2013a} for single 
impurity problems, we reduce a complex lattice geometry to a single chain, or a 
multi-leg ladder in the case of multiple impurities.

\begin{figure}
\centering
\includegraphics[width=0.43\textwidth]{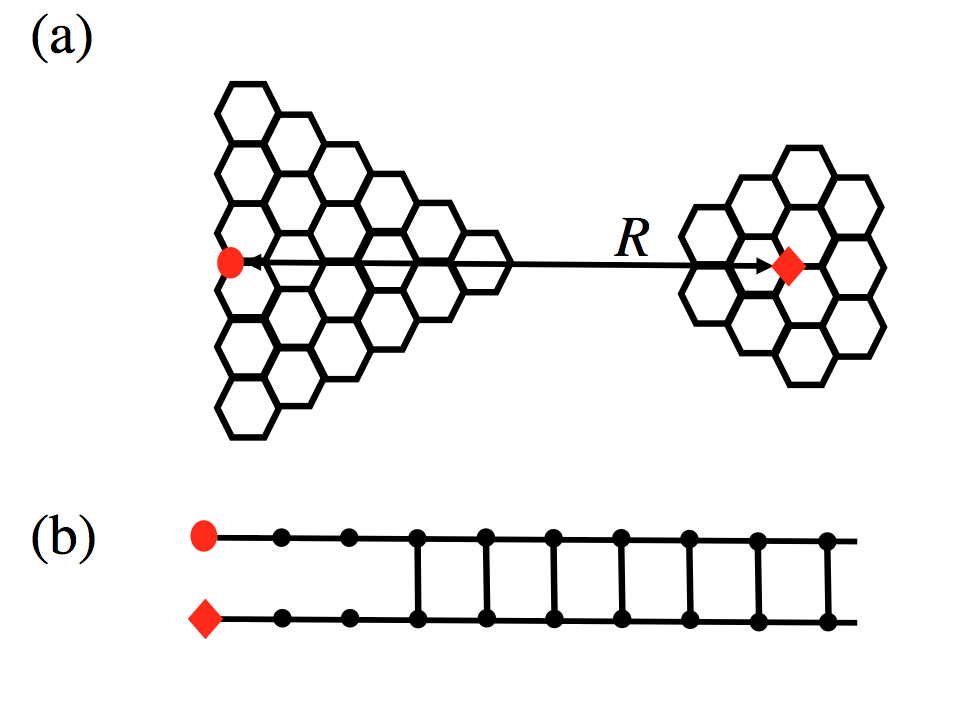}
\caption{(a) Sketch of the graphene geometry studied through the Lanczos transformation for one impurity sitting at the edge (circle) and the measurement site in the bulk at a distance $R$ (diamond). In (b) we show the geometry of the equivalent problem, with the two seed orbitals coupled to non-interacting tight-binding chains. The Lanczos orbitals only start interfering at a distance $R/2$, introducing a hopping term between the chains. There is an exact canonical transformation connecting the two problems.
}
\label{figure1}
\end{figure}

In order to measure spin-spin correlations in real space, we need to employ the 
multi-impurity formulation described in detail in Refs.~\onlinecite{Allerdt2015,Yunoki.Block}. In 
this approach, we use two seed states corresponding to the impurity site and the orbital 
where the correlations will be measured. A block Lanczos recursion will generate a block tridiagonal
matrix that can be interpreted as a single-particle Hamiltonian on a ladder geometry. This is illustrated 
schematically in Fig.~\ref{figure1}(a), showing the case of one seed state at the edge, 
and the second one somewhere in the bulk. The typical equivalent problem we need to solve 
numerically is depicted in the panel Fig.~\ref{figure1}(b). 
Due to the presence of spin-orbit coupling in the bulk, this geometry cannot be further 
simplified. Still, the Hamiltonian in the new basis will be one-dimensional, and local, 
{\it i.e.}, the many-body terms are still the same as in the original 
Anderson impurity coupled to the original lattice.

We want to emphasize that this mapping (see Fig.~\ref{figure1}) 
is exact and both geometries are connected by a unitary transformation. 
The combination with the DMRG method 
allows us to obtain exact results with \emph{real space resolution} and uncover the 
marked difference between crest and trough sites, as 
described above, which is inaccessible to the majority of other methods traditionally used to study magnetic 
impurity models. 
Results in this work were obtained by keeping up to 3000 DMRG states, 
which grants an accuracy of the order of $10^{-6}$ or better for the energy and correlations.

\begin{figure}
\centering
\begin{minipage}{0.23\textwidth}
\centering
\includegraphics[width=\textwidth]{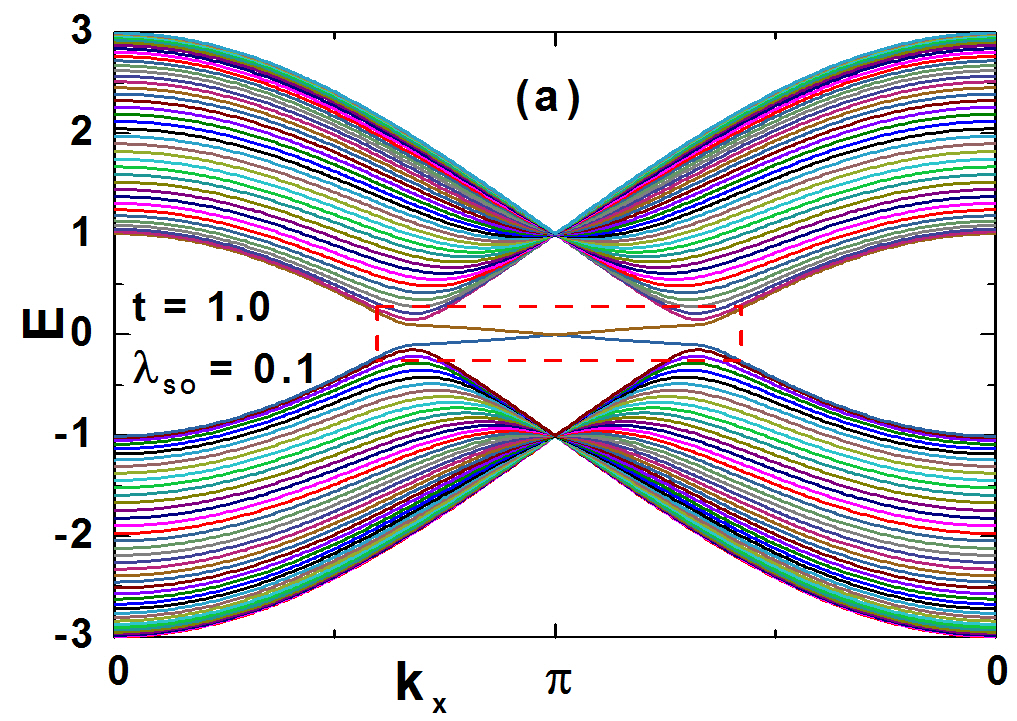}
\end{minipage}
\hfill
\begin{minipage}{0.23\textwidth}
\centering
\includegraphics[width=\textwidth]{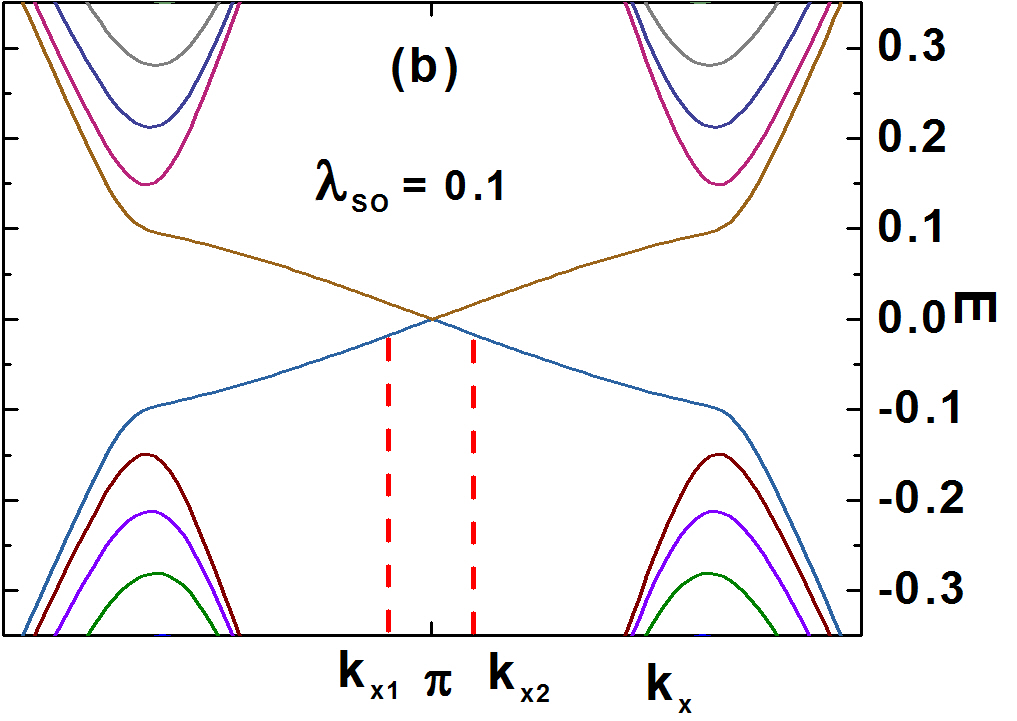}
\end{minipage}
\centering
\begin{minipage}{0.23\textwidth}
\includegraphics[width=\textwidth]{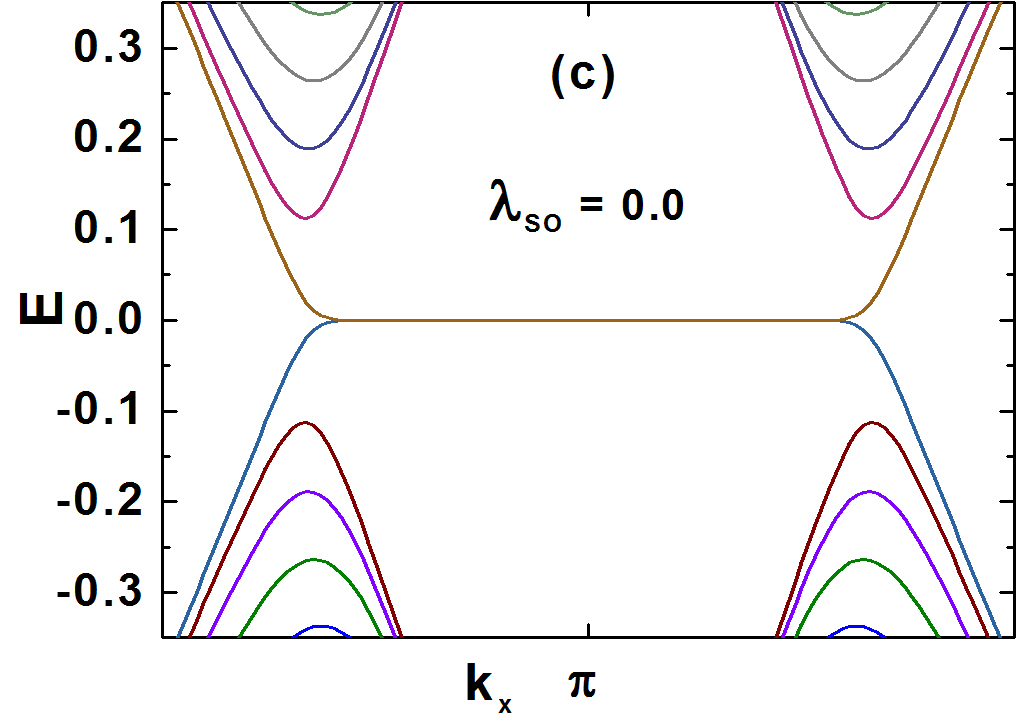}
\end{minipage}
\hfill
\begin{minipage}{0.23\textwidth}
\centering
\includegraphics[width=\textwidth]{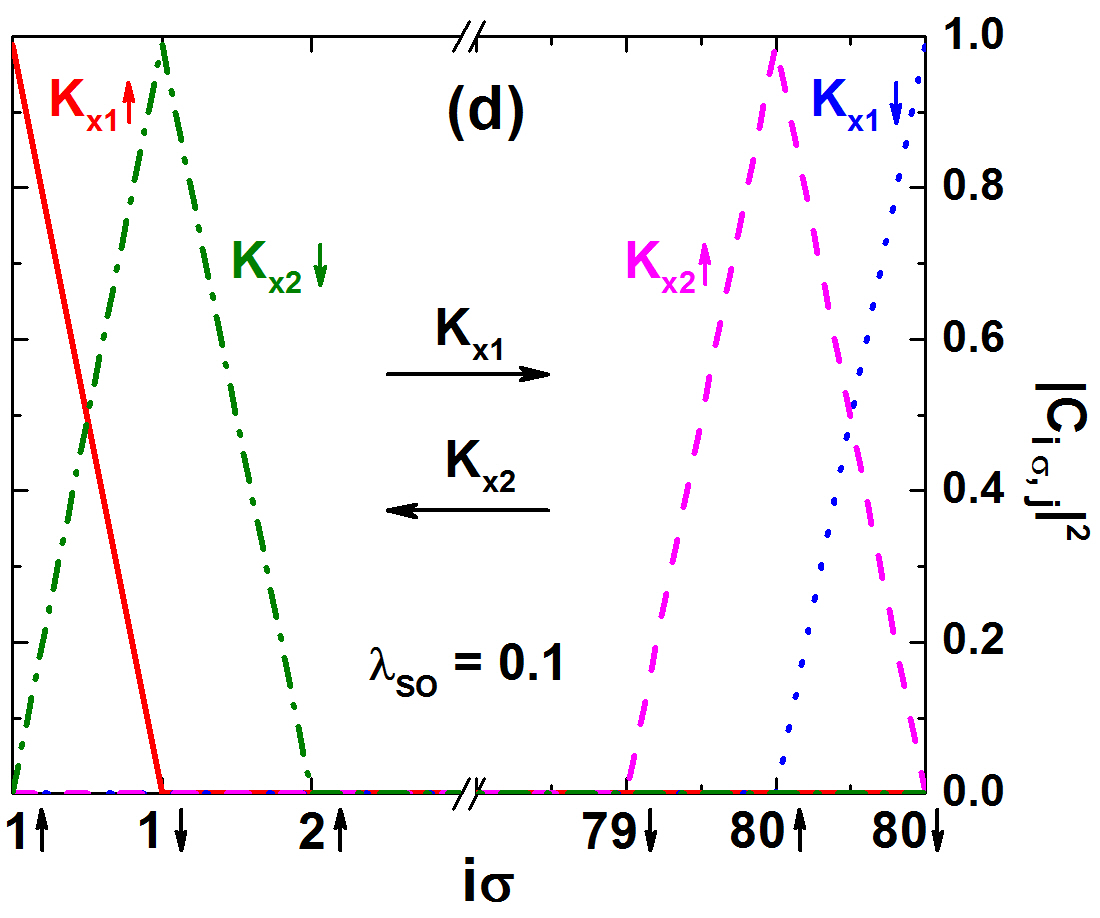}
\end{minipage}
\caption{(Color online) (a) The energy spectra of a zigzag stanene nanoribbon for $\lambda_\text{SO}=0.1$. 
	(b) Zoom on the (red) dashed square in panel (a) showing details of the energy 
	dispersion of the edge states. (c) Similar to panel (b), but for $\lambda_\text{SO}=0.0$, 
	which is appropriate for graphene. (d) Value of $|C_{i\sigma}|^2$ for the four metallic 
	edge states associated to the wave vectors $k_{x1}$ and $k_{x2}$ in panel (b). Note that, 
	as discussed in more detail in the text, they are located exclusively at crest sites, 
	and states with different spin polarizations propagate in opposite directions at each edge. 
}
\label{figure2}
\end{figure}

\begin{figure}
\centering
\begin{minipage}{0.23\textwidth}
\centering
\includegraphics[width =\textwidth]{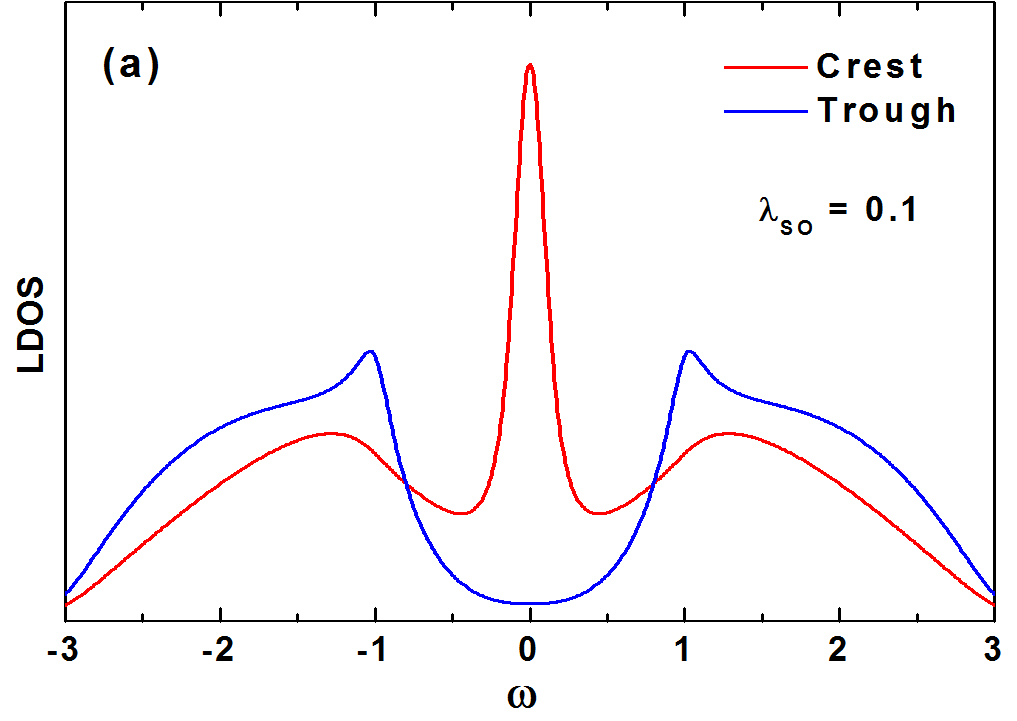}
\end{minipage}
\hfill
\begin{minipage}{0.23\textwidth}
\centering
\includegraphics[width =\textwidth]{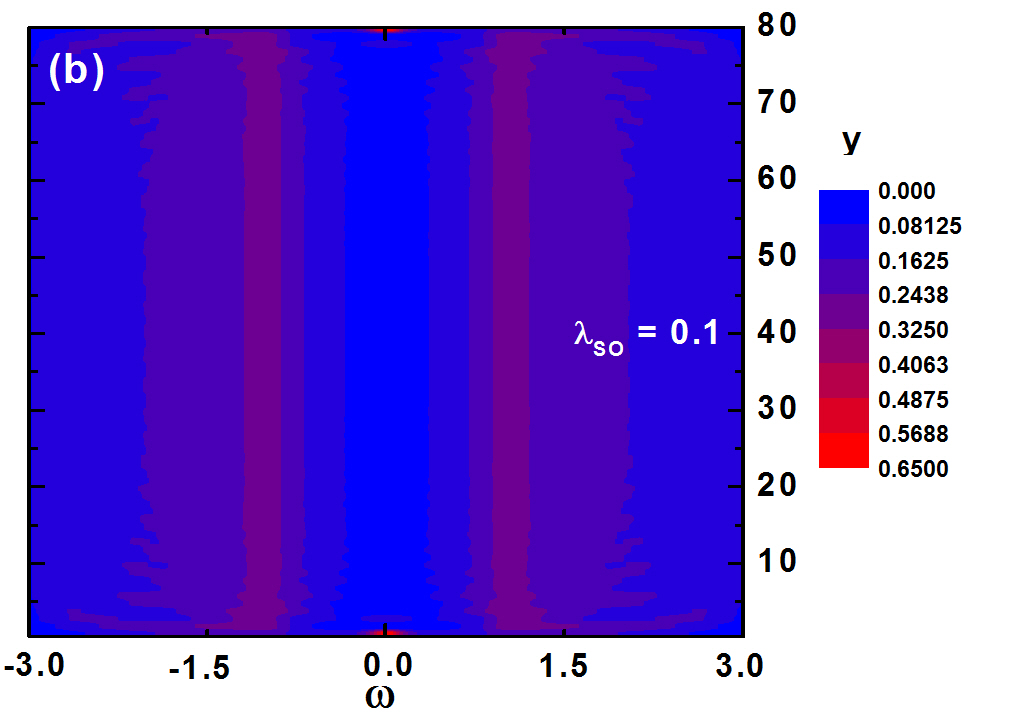}
\end{minipage}
\caption{(Color online) (a) LDOS for two edge sites in the TI phase ($\lambda_\text{SO}=0.1$): 
the crest site LDOS [dark (red) curve] 
displays a pronounced spectral density peak at the Fermi energy, while the trough site LDOS 
[gray (blue) curve] has very small, but finite spectral density at the Fermi energy. (b) Color contour plot 
showing the LDOS across the width of a nanoribbon $N=80$ sites wide ($\lambda_\text{SO}=0.1$). Notice how the spectral 
density at the Fermi energy is located mostly at the edges [and primarily at crest sites, 
as shown in panel (a)], while the bulk remains gapped. 
}
\label{figure3}
\end{figure}

\section{Results}
\subsection{Zigzag nanoribbon band structure}\label{ribbon-band}

Fig.~\ref{figure2}(a) shows the band structure obtained for 
$H_{{\rm band}}$ in a ZNRB with periodic boundary conditions (PBC) in the $x$-direction and 
open boundary conditions (OBC) in the $y$-direction. The SOI used was $\lambda_\text{SO}=0.1$  
and the nanoribbon is $80$ sites across in the OBC direction. Each band is double degenerate, 
besides the Kramers $\omega(k_x)=\omega(-k_x)$ degeneracy, and the energy spectrum is particle-hole symmetric 
around $\omega=0$ (half-filling). In panel (b) we show a zoom into the dashed (red) box shown in panel (a), 
focusing on the bands that straddle across the valence and conduction bands. 
These bands host the 4 helical edge states (two in each edge). 
Panel (c) shows the same zoom, but for $\lambda_\text{SO}=0.0$ (appropriate for graphene), 
showing the well studied flat bands associated to the edge states present in a graphene ZNRB (see 
Ref.~\onlinecite{CastroNeto2006a} for details).

In Fig.~\ref{figure2}(d) we see a plot of the coefficient squared, for each site across 
the ZNRB, for the four edge states that are 
located symmetrically around $k_x=\pi$, {\it i.e.}, $k_{x1}=\pi - \delta$ and $k_{x2}=\pi + \delta \equiv -2\pi+k_{x2}=-k_{x1}$. 
Note that $k_{x1}$ and $k_{x2}$ are time reversed momenta (Kramers related), indicating propagation in opposite directions. 
The location of $k_{x1}$ and $k_{x2}$, and the associated edge states, are schematically 
indicated in panel (b). The results are for $\lambda_\text{SO}=0.1$  and $\delta=\pi/64$. 
For a ZNRB with width of $N=80$ sites, each state in the 160 bands (including spin) for an arbitrary value of $k_x$ 
can be written as
\begin{eqnarray}
	|E_j(k_x)\rangle = \sum_{i,\sigma} C_{i\sigma,j}(k_x) |i\sigma\rangle
\end{eqnarray}
where $i=1,\cdots,80$ runs through the sites across the OBC direction, $\sigma=\uparrow\downarrow$ indicates spin, 
$E_j(k_x)$ (for $j=1,\cdots,160$) runs through all the eigen-energies for a specific $k_x$ value, 
and $|i\sigma\rangle$ is a localized orbital state on site $i$ with spin $\sigma$. 
Taking $k_x=k_{x1},k_{x2}$ as defined above and 
$j$ values corresponding to the double-degenerate bands connecting the valence and conduction bands 
($j=79$ to $82$, {\it i.e.}, the 4 edge states), we plot $|C_{i\sigma,j}(k_x)|^2$ for each site $i\sigma$. 
In reality, we plot only the values for the edge sites (crest sites at each edge), since all the other coefficients vanish 
($|C_{1\uparrow}|^2=0.99017$, for example). As indicated by the labels in Fig.~\ref{figure2}(d) 
($k_{x1}$$\uparrow$ and $k_{x2}$$\downarrow$), 
states in the $i=1$ edge propagate in opposite directions, with opposite spin polarizations, and the reverse occurs 
($k_{x1}$$\downarrow$ and $k_{x2}$$\uparrow$) on the $i=80$ edge. 
Therefore, these helical edge states are very localized, propagating through what we dubbed as crest sites, 
presenting the characteristic locking of spin and 
momentum that is associated to the non-trivial topological phase created by the SOI \cite{Hasan2010b}. 
In addition, narrower ZNRB have less localized states, the localization varying slightly with $k_x$ as we 
move away from $k_x=\pi$, and the states become (discontinuously) completely delocalized 
(becoming bulk states) once these two edge bands merge with either the conduction or the valence band. 

To close this section, we make a few remarks on the effects of electronic interactions. These are typically too 
weak to introduce noticeable spin correlations in the bulk, especially due to the vanishing density of states 
at the Dirac points. However, theory has predicted that electronic repulsion causes a hybridization between 
edge states on opposite sides of a ZNRB, inducing edge ferromagnetism 
\cite{Son2006,Son2006a,Yazyev2008,Tao2011,Schmidt2013,Jin2014}. These effects depend on both 
the magnitude of the interactions and the width $W$ of the ribbon, and decay very rapidly\cite{Jung2009} 
as $W^{-2}$. 

An analysis of the literature indicates that 
there is a great deal of controversy regarding the experimental verification of the theoretical prediction 
of ferromagnetic edge states at the edges of a ZNRB \cite{Makarova2015}: for example, all the 
measurements intended to uncover them are constrained 
to either charge transport or scanning tunneling microscopy experiments, whose results are open to alternative interpretations. 
Therefore, the current consensus seems to be that the existence of the magnetic edge states 
has not been settled yet.  Clearly, magnetism would break time reversal symmetry,
compromising the existence of the topological edge states.  
Recent work by 
Lado and Fern\'andez-Rossier \cite{Lado2014} shows, through mean-field calculations, that a large 
enough SOI can suppress the magnetic moment of the edge states. Nonetheless, this problem falls beyond the scope of our work, which focuses on understanding  
the properties of the Kondo state in the absence of magnetism.

\begin{figure}
\centering
\includegraphics[width =0.4\textwidth]{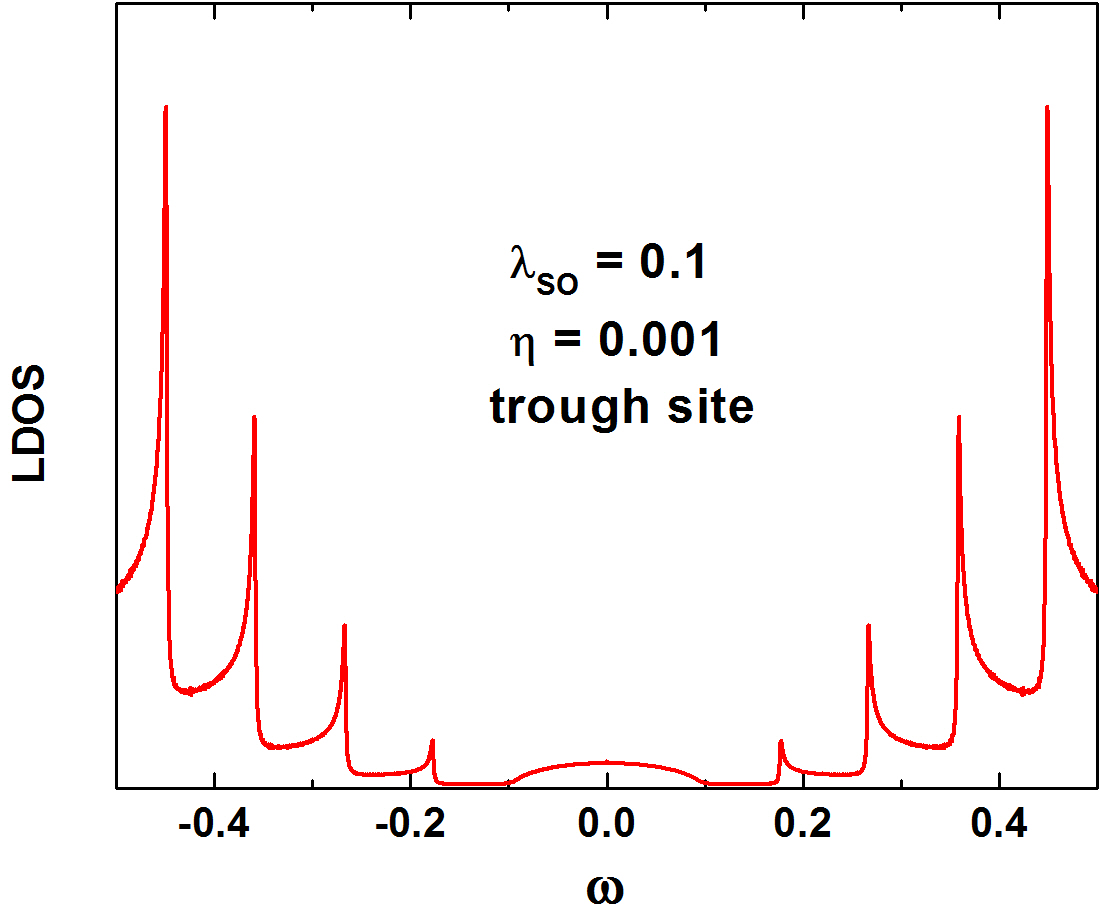}
\caption{Detail of the LDOS at a trough site, using a large number of poles 
	and a small broadening $\eta=0.001$ of the Green's function imaginary part. 
	The spectral weight in the pesudogap-like 
	region around the Fermi level is small but finite \cite{note2}.
}
\label{figure4}
\end{figure}

\subsection{Zigzag nanoribbon local density of states}\label{ldos}

Figure \ref{figure3}(a) shows the LDOS for a crest [(red) dark gray curve] and for a trough [(blue) light gray] 
site at the edge of the ZNRB. In agreement with the results shown 
in Fig.~1(d), the conducting edge states [with spectral density at the Fermi 
energy ($\epsilon_F=0$)] are mostly located at crest sites. Panel (b) shows a contour plot 
of the LDOS for all sites in the OBC direction. The spectral density 
at the Fermi energy is restricted to the edge [primarily to a crest site, as shown in panel (a)] while 
all the sites away from the edge present an insulating spectra, where the size of the gap 
partially derives from the spin-orbit interaction and mostly from the confinement along 
the OBC direction (the wider the nanoribbon, the smaller is the energy gap). It is important to notice that, 
unlike the insulating bulk, the LDOS at the trough sites displays a small but finite LDOS at 
the Fermi energy (for a detailed view, with increased resolution, see Fig.~\ref{figure4}), leaving 
room for the formation of a Kondo state. 

\begin{figure}
\centering
\includegraphics[width = 0.4\textwidth]{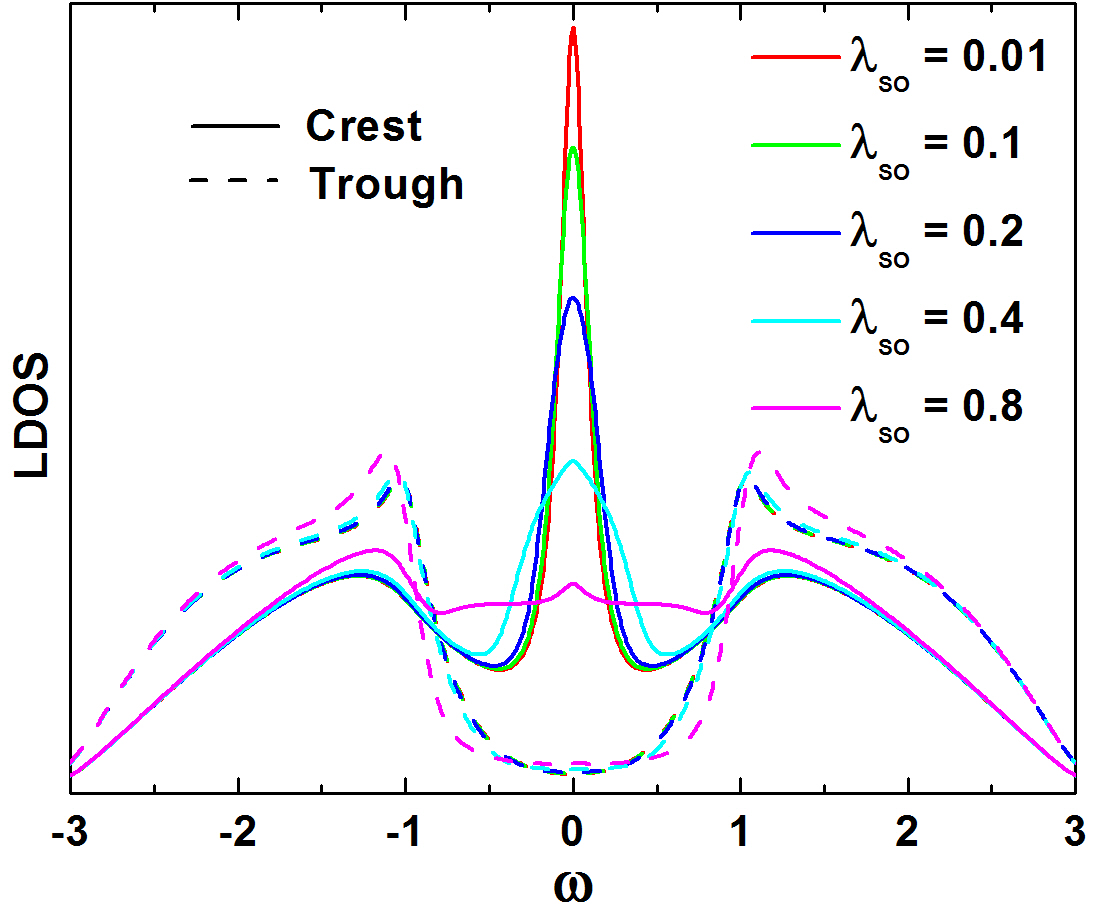}
\caption{(Color online) Variation of the LDOS with the value of $\lambda_\text{SO}$ 
for a crest (solid lines) and a trough (dashed lines) site. 
By varying $\lambda_\text{SO}$ by almost two orders of magnitude, from $10^{-2}$ (most pronounced 
peak at $\omega=0$) to $0.8$ (almost flat at $\omega=0$), we see that the main change 
	in the LDOS of a crest site is the continuous spread of the 
spectral weight (located at the Fermi energy peak) into an almost flat distribution 
that covers the gap. The LDOS of a trough site almost does not vary with $\lambda_\text{SO}$. 
}
\label{figure5}
\end{figure}

In Fig.~\ref{figure5} we report the effects of SOI in the LDOS. 
By increasing $\lambda_\text{SO}$ from $0.01$ to $0.8$ the spectral weight peak at the 
Fermi energy gradually broadens until a flat distribution closes the gap. The 
results for a trough site (dashed curves) show that aside from a slight increase 
of the LDOS at the edge of the gap and a slight increase of the gap itself, 
the LDOS changes only marginally. 
A contour plot for the LDOS across the nanoribbon for $\lambda_\text{SO}=0.8$ (not shown), similar to Fig.~\ref{figure3}(a), 
shows that the bulk sites do not qualitatively change their LDOS, {\it i.e.}, the gap in the bulk remains intact. 

\subsection{Spin correlations}\label{sscorr}

\begin{figure}
\centering
\begin{minipage}{0.48\textwidth}
\centering
\includegraphics[width=0.75\textwidth]{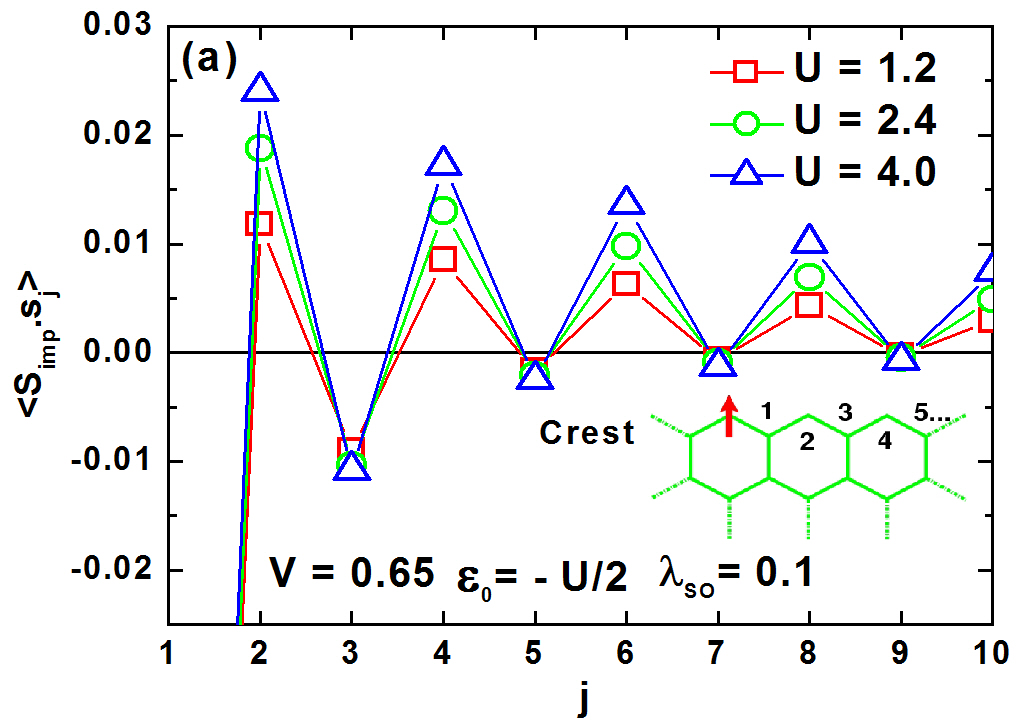}
\end{minipage}
\vfill
\begin{minipage}{0.48\textwidth}
\centering
\includegraphics[width=0.75\textwidth]{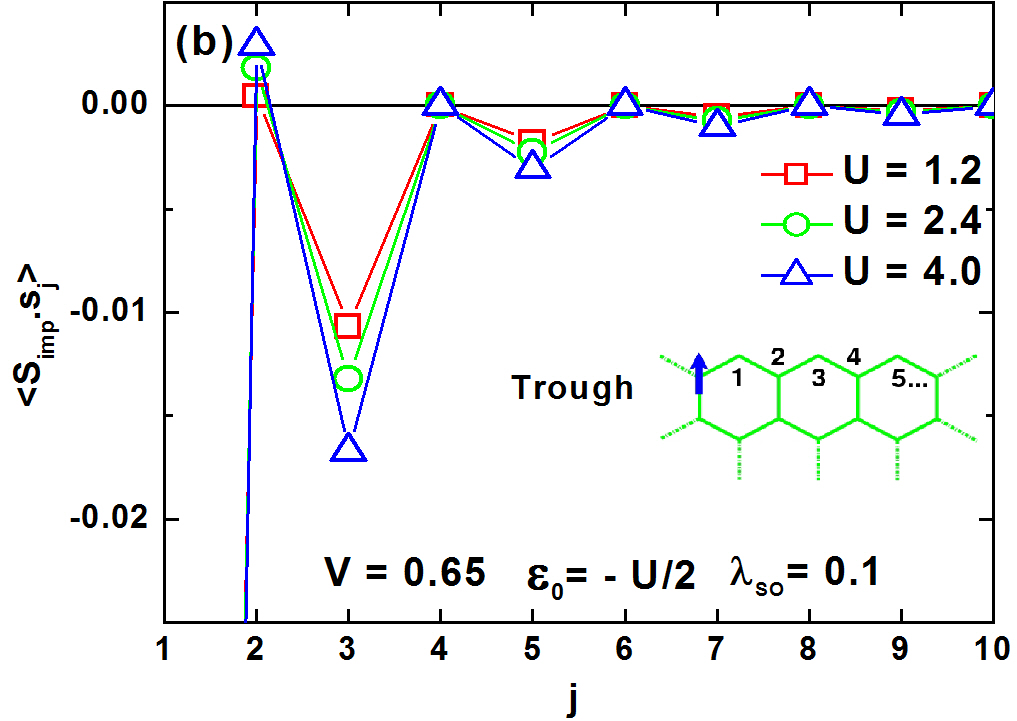}
\end{minipage}
\caption{(Color online) Spin correlations for a substitutional 
impurity located at a crest (a) and trough (b) edge site, respectively. 
Results show correlations along edge sites $j$ for $U=1.2$ [(red) squares], $U=2.4$ 
[(green) circles], and $U=4.0$ [(blue) triangles], $V=0.65$, and $\lambda_\text{SO}=0.1$. 
}
\label{figure6}
\end{figure}

\begin{figure}
\centering
\includegraphics[width =0.38\textwidth]{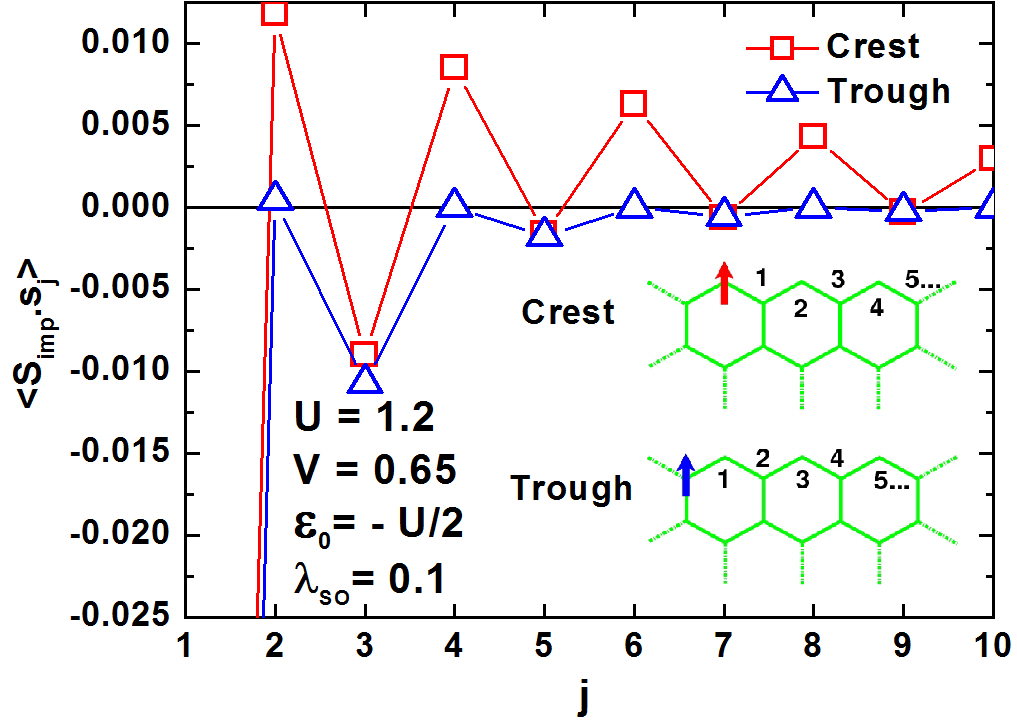}
\caption{(Color online) Comparison of spin correlations between crest [(red) squares] 
and trough [(blue) triangles] sites for a substitutional impurity. The parameter values are $V=0.65$, $\lambda_\text{SO}=0.1$, 
and $U=1.2$. There is a clear qualitative difference between 
impurities sitting at crest and trough sites. 
}
\label{figure7}
\end{figure}

Our model takes into account a real-space description of the lattice. Due to the spatial dependence of 
the density of states, the physics of the magnetic impurity will vary accordingly, displaying 
important differences determined by its location. This will be clearly visible in the spin correlations results. 
Fig.~\ref{figure6} shows the spin correlations between the impurity and the surrounding conduction electrons for a 
substitutional impurity (replacing a silicon atom) sitting at crest and trough sites in panels 
(a) and (b), respectively. The spin correlations $\langle \vec{S}_{\mathrm{imp}} \cdot \vec{s}_j \rangle$  
are calculated for sites $j$ along the edge, as indicated.
Results correspond to three different values of $U=1.2$ [(red) squares], $2.4$ [(green) 
circles], and $4.0$ [(blue) triangles]. 
Correlations are strong at short distances, their magnitude falls away from the 
impurity, and their range increases with $U$. In addition, 
in both cases, correlations with sites in the opposite sublattice to the one where the impurity 
is located (odd sites) are antiferromagnetic, while they are ferromagnetic for same sublattice 
sites (even sites).
These results are typical of those obtained for the Kondo effect for an $S=1/2$ impurity 
connected to a non-interacting chain (see, for example, Ref.~\onlinecite{Nuss2015}). Ferromagnetic correlations dominate for crest site impurities, as opposed to trough site impurities, where antiferromagnetic correlations dominate. 
In addition, crest site correlations have larger magnitude and are longer ranged than those for trough sites. 

\begin{figure}
\centering
\includegraphics[width = 0.38\textwidth]{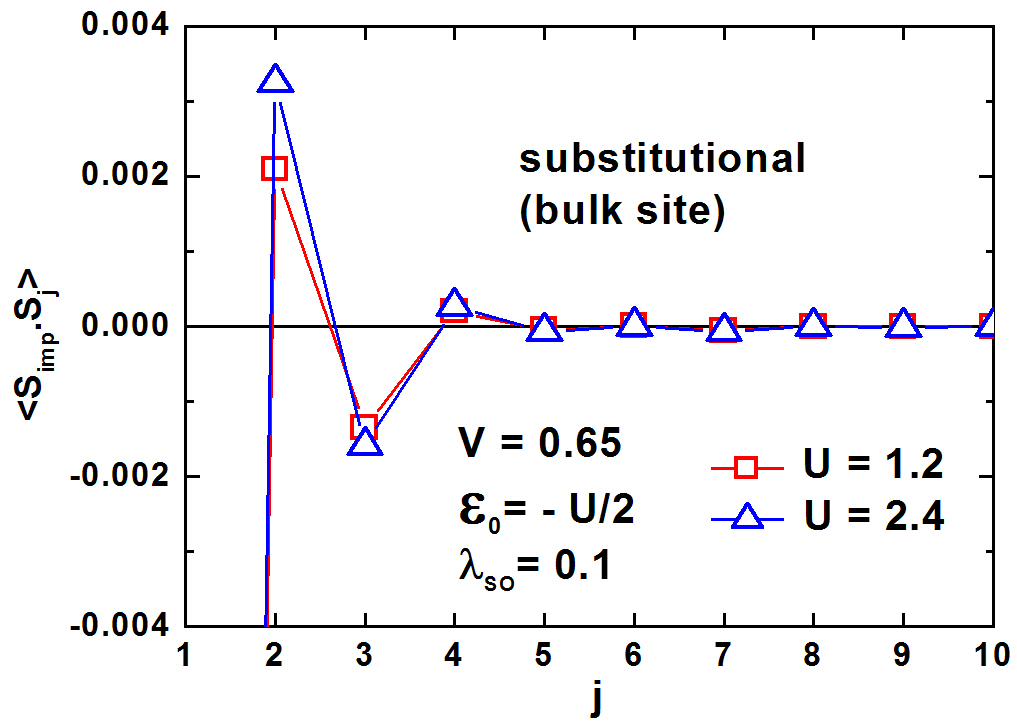}
\caption{(Color online) Spin correlations for a substitutional impurity located five lattice spacings,
counting from the edge, into the bulk, for $U=1.2$ [(red) squares], $U=2.4$ [(blue) triangles], 
and $\lambda_\text{SO}=0.1$. 
The impurity forms a localized singlet with the three neighboring spins, differently from a trough site, 
	whose correlations extend along the edge [see Fig.~\ref{figure6}(b)]. 
	Note that the line of $j$-sites probed is parallel to the edge.
}
\label{figure8}
\end{figure}

It is also interesting to note that an impurity located either in a crest or a trough site correlates much more strongly 
to crest sites along the edge. In contrast, the dominant (antiferromagnetic) correlations for a trough site impurity 
have very short range and are essentially independent of $U$, 
while the dominant (ferromagnetic) correlations  for a crest site impurity decay slowly and increase slightly with $U$. 
The presence of ferro or antiferromagnetic correlations stems naturally from the expected structure of the 
RKKY interaction on a bi-partite lattice \cite{Saremi2007,brey2007,Allerdt2017}. However, the vanishing of the correlations at 
even distances from the trough sites can be attributed to the nodal structure of the electronic 
wavefunctions near the Fermi level, that interfere destructively to yield a very small amplitude on 
those sites \cite{Allerdt2015}. 

As seen in Fig.~\ref{figure7},  
impurities at crest sites induce large and slowly decaying hybridization clouds 
with dominant ferromagnetic correlations that decay algebraically with distance 
[(red) squares in Fig.~\ref{figure7}]. This can be understood in terms of crest sites forming an 
effective one-dimensional channel \cite{Wu2006,Xu2006}.
 On the other hand, impurities at trough sites form very small clouds 
[(blue) triangles in Fig.~\ref{figure7}], as if sitting in small metallic ``puddles'' 
that leak from the metallic edges, surrounded by the bulk. 
To illustrate that the results for a trough site are qualitatively different from those 
for a bulk site, Fig.~\ref{figure8} shows the correlations around a 
substitutional impurity in the bulk, five lattice spacings 
from the edge (note that the vertical axis scale in Fig.~\ref{figure8} is considerably 
smaller than that in Fig.~\ref{figure7}). When located in a bulk site, the impurity forms a localized singlet 
with its three neighbors, completely decoupled from the bulk. Correlations in 
this case are practically identical to the two-site problem with an Anderson impurity 
connected to a single non-interacting site (not shown). 

\begin{figure}
\centering
\begin{minipage}{0.48\textwidth}
\centering
\includegraphics[width =0.75\textwidth]{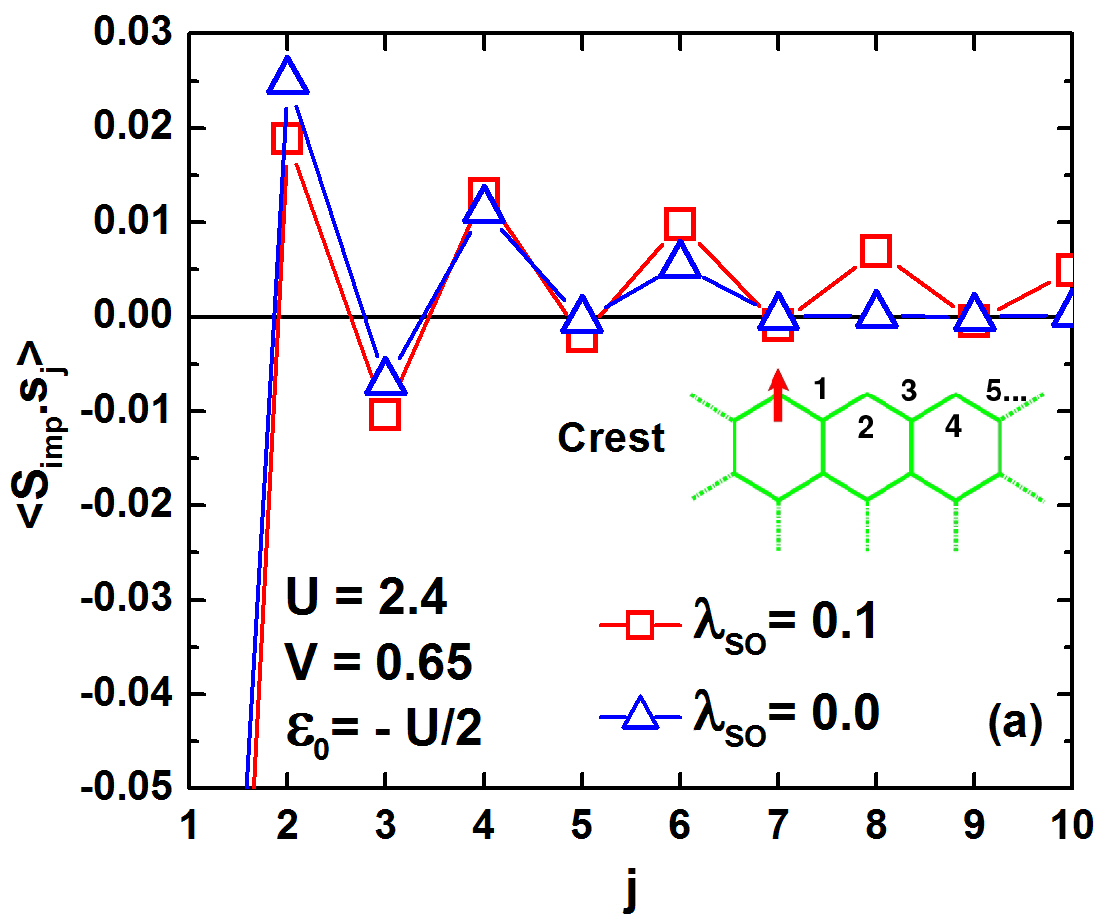}
\end{minipage}
\vfill
\begin{minipage}{0.48\textwidth}
\centering
\includegraphics[width =0.75\textwidth]{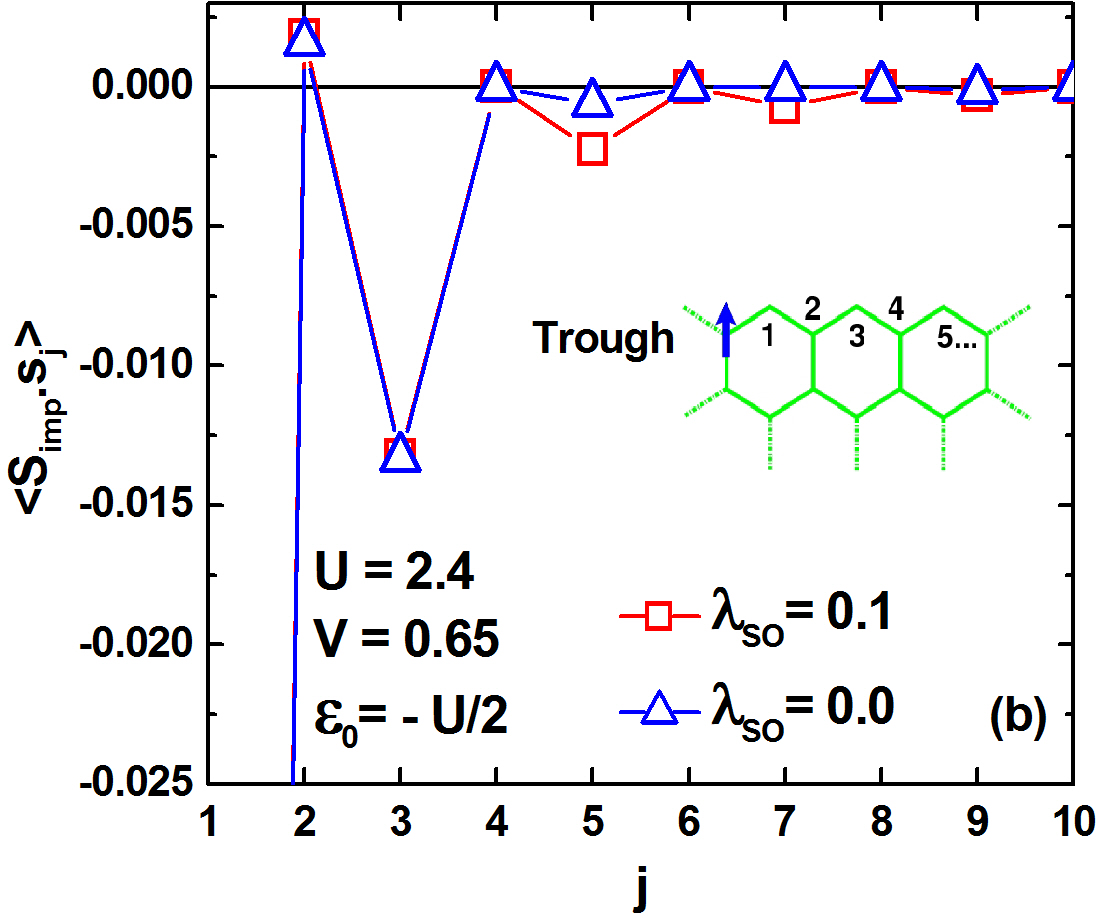}
\end{minipage}
\caption{(Color online) Effect of SOI on spin correlations for  
substitutional impurity at crest (a) and trough (b) sites. Results 
are for $U=2.4$. The SOI causes an overall increase of the spin correlations. 
}
\label{figure9}
\end{figure}

\begin{figure}
\centering
\begin{minipage}{0.48\textwidth}
\centering
\includegraphics[width =0.75\textwidth]{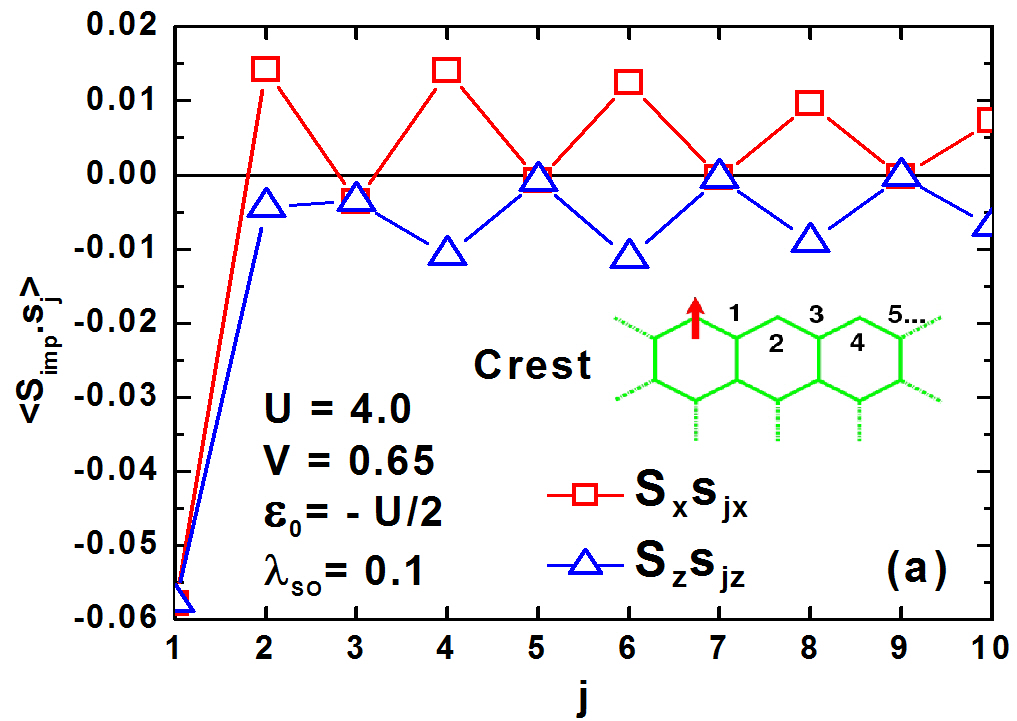}
\end{minipage}
\vfill
\begin{minipage}{0.48\textwidth}
\centering
\includegraphics[width =0.75\textwidth]{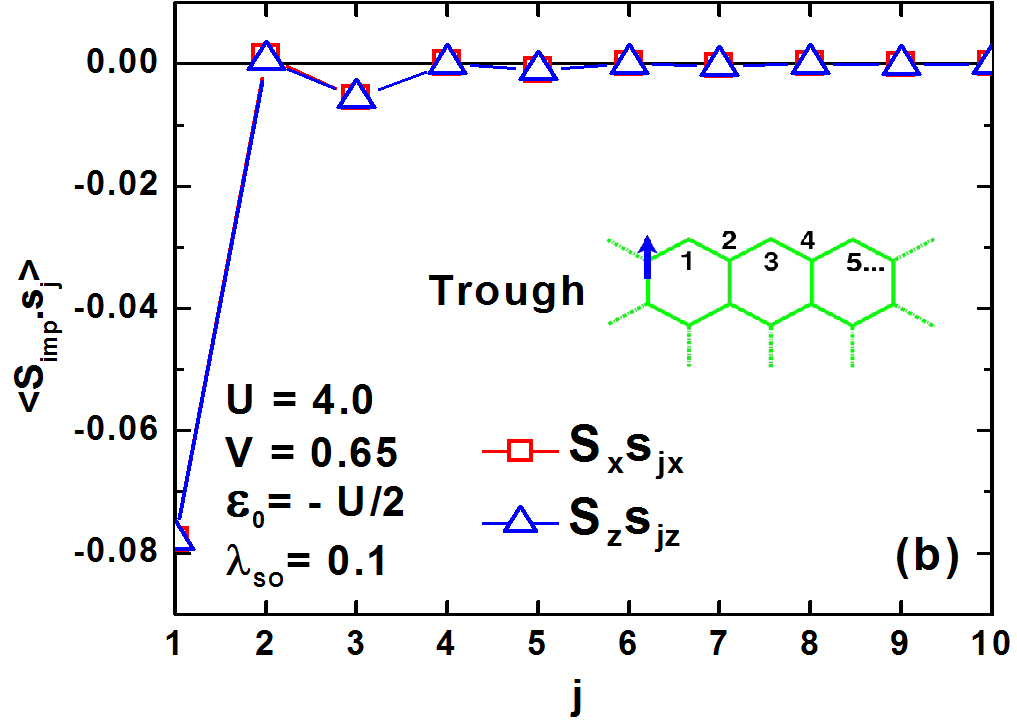}
\end{minipage}
	\caption{(Color online) Comparison between $\langle S_{\mathrm{imp}x}s_{jx} \rangle$ 
	and $\langle S_{\mathrm{imp}z}s_{jz} \rangle$ for 
a substitutional impurity at a crest (a) and a trough (b) site for $U=4.0$, $V=0.65$, and $\lambda_\text{SO}=0.1$. 
The anisotropy is expected as a consequence of the spin-orbit interaction \cite{Zarea2012,Vernek2016}. 
As can be seen by comparing panels (a) and (b), the anisotropy is much larger for a crest site and it increases with $U$ 
	[see Fig.~\ref{figure6}(a)]. 
}
\label{figure10}
\end{figure}

We proceed to compare spin correlations with and without SOI to understand 
how it changes the coupling of a magnetic impurity to the edge states. Fig.~\ref{figure9}
shows results for a substitutional impurity, at both crest and trough sites, 
demonstrating an overall increase of the correlations when the SOI is introduced. 
At first sight, these results seem to indicate that the SOI does not change 
qualitatively the results. However, it is important to notice that the SOI induces an anisotropy 
in the spin correlations ($\langle S_zs_{jz} \rangle \neq \langle S_xs_{jx} \rangle$) 
that stems from an emergent Dzyaloshinskii-Moriya interaction between the impurity and the conduction 
electrons \cite{Zarea2012,Vernek2016}, as illustrated in Fig.~\ref{figure10}. This effect is clearly visible for 
the case of impurities sitting at crest sites, with transverse $XY$ correlations  [(red) squares] and along the 
$Z$ direction [(blue) circles] having opposite sign, indicating helical order. However, for impurities located at trough 
sites [Fig.~\ref{figure10}(b)] this effect is very small or practically nonexistent.

\begin{figure}
\centering
\includegraphics[width =0.48\textwidth]{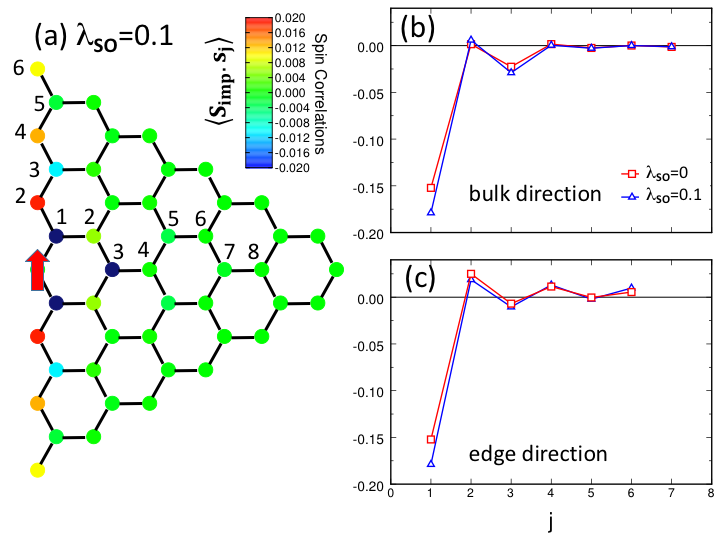}
\caption{(Color online)  
	(a) Color map of the spin correlations for a substitutional impurity at a crest site, for $U=2.4$, 
	$V=0.65$, and $\lambda_\text{SO}=0.1$. 
	(b) and (c) show results along the edge and perpendicular to it, respectively, for 
	$\lambda_\text{SO}= 0.0$ [(red) squares] and $\lambda_\text{SO}=0.1$ [(blue) triangles]. 
	The actual positions of the $j$ sites in (b) and (c) are indicated in panel (a).
}
\label{figure11}
\end{figure}

As demonstrated above, the spatial resolution 
of our calculations are instrumental at determining the full structure of the spin correlations 
and uncover the response of the edge states to the 
presence of substitutional magnetic impurities. 
Results for a crest-site impurity are summarized in Fig.~\ref{figure11}. 
Panel (a) presents a color map of the impurity spin correlations in an extended region surrounding the substitutional atom.  
 Details along the bulk and edge directions are shown in panel (b) and (c), respectively.  
The actual position of the $j$ sites for panels (b) and (c) are indicated in 
 (a). From these results it emerges that the edge state is scattered and ``snakes'', or circles
around the impurity\cite{Wu2006}. 
This effect implies that the problem may not be trivially studied using a one-dimensional 
lead to represent the edge. This was already pointed out through a no-go theorem in Ref.~\onlinecite{Wu2006} 
stating that a helical liquid with an odd number of modes cannot emerge from a 
purely one-dimensional model. In addition, even though the impurity is substitutional, 
it appears as though it is {\it side coupled} to the edge \cite{Vernek2016}, and not embedded in it.

\subsection{Kondo temperature}\label{stk}
Anderson showed~\cite{Anderson1970}, with his ``Poor man's scaling'', that the Kondo problem can 
be treated perturbatively at energies larger than the so-called Kondo temperature $T_K$, which is 
the only relevant energy scale, and does not depend on the high energy details.
A renormalization group analysis shows that the system flows toward an attractive strong coupling 
fixed point, described by a tightly bound state formed by the impurity and the conduction 
electrons, the ``Kondo singlet''. In this regime one can show that many quantities satisfy 
a universal scaling characterized precisely by $T_K$. This quantity has a strict 
universal meaning in the thermodynamic limit (or rather, in the universal scaling regime). In 
finite systems, like the one we presently discuss, one can define a similar energy scale 
as the energy gained by the system by forming a Kondo singlet, the correlation energy: 
\begin{equation}
E_\mathrm{corr} = E_0 - E_\mathrm{proj},
\end{equation}
where $E_0$ is the ground state energy, and 
\begin{equation}
	E_\mathrm{proj} = \frac{\langle \mathrm{g.s.} |S^-_{imp} H S^+_{imp} | \mathrm{g.s.} \rangle}
	{\langle \mathrm{g.s.} |S^-_{imp} S^+_{imp} | \mathrm{g.s.}\rangle}, 
\end{equation}
with $| \mathrm{g.s.} \rangle$ being the ground state. 
The operators $S^\pm_{imp}$ act on the impurity site and project the ground state singlet 
onto a 
state where the impurity and the bulk are disentangled, thus forming a product state. This 
clearly is a variational estimate of the correlation energy, and comparisons to the dynamical spin correlations 
show that indeed it is an accurate measure of $T_K$ \cite{note3}. Notice that, even though the 
Hamiltonian is gapless, $E_\mathrm{corr}$ is finite. 

\begin{figure}
\centering
\includegraphics[width =0.48\textwidth]{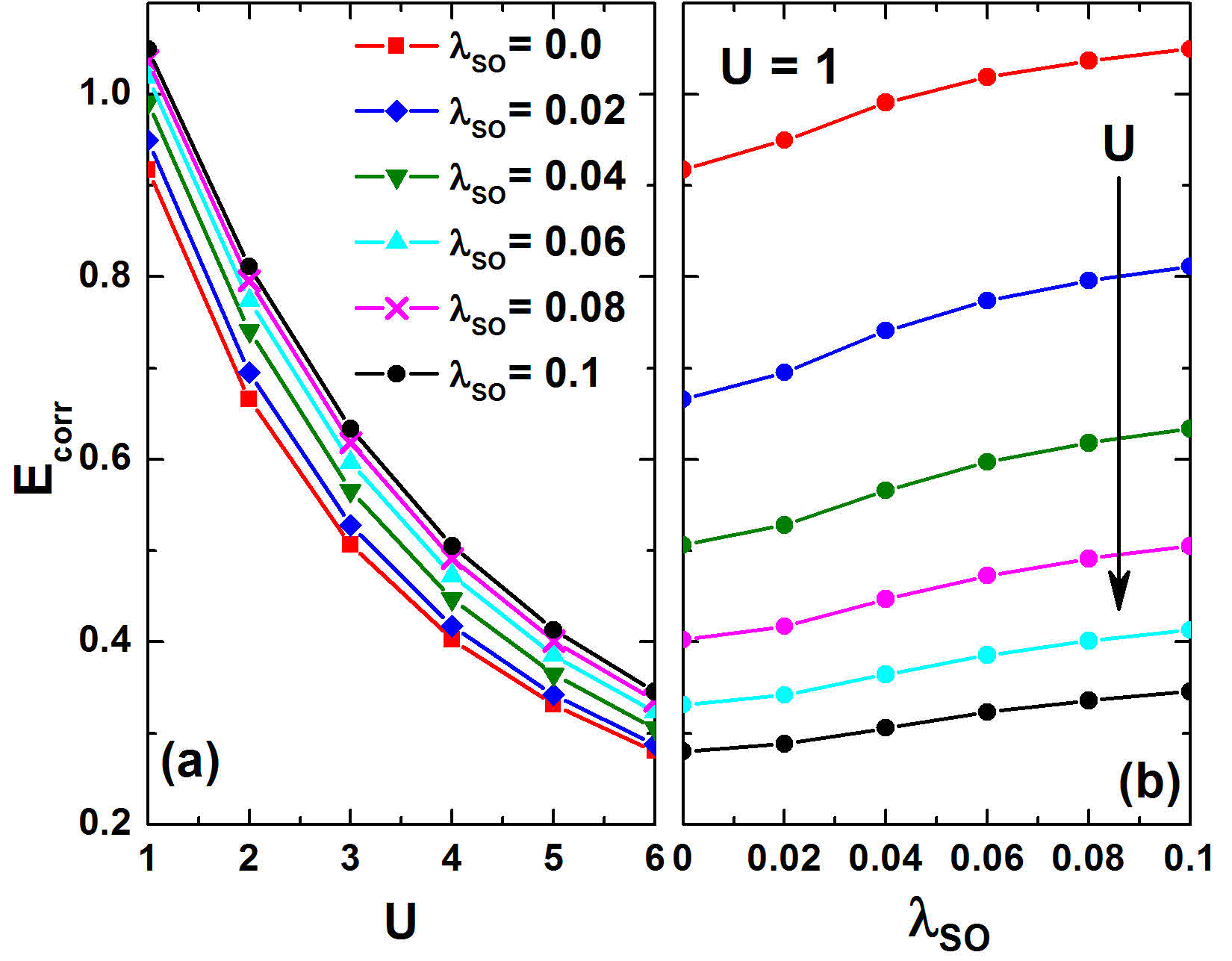}
\caption{(Color online) Characteristic energy scale $E_\mathrm{corr}$ (see text for definition), which is a measure of the 
	Kondo temperature, as a function of (a) the Coulomb interaction $U$ and (b) the spin 
	orbit coupling $\lambda_{\text{SO}}$. Results in (b) are for the same values of $U$ 
	as in (a), and the strength of the interaction increases as indicated by the arrow.}
\label{figure12}
\end{figure}

Our results for $E_\mathrm{corr}$, for an impurity at a crest site, are shown in 
Fig.~\ref{figure12}(a) as a function of 
the interaction $U$ for different values of SOI ($0.0 \leq \lambda_{\text{SO}} \leq 0.1$), and in  
panel (b) as a function of $\lambda_{\text{SO}}$ for different 
values of interaction $1 \leq U \leq 6$ (varying in steps of 1). 
Unlike prior work by Zarea {\it et at.} \cite{Zarea2012} that predicts an exponential enhancement of 
the Kondo temperature in the presence of SOI, we find that this enhancement is very weak 
[see panel (b)], 
in agreement with numerical renormalization group treatments of this problem~\cite{Zitko2011,Mastrogiuseppe2014}.

\section{Summary and Conclusions}\label{conclusion}
We have applied the Lanczos transformation method combined with the DMRG \cite{Busser2013a,Allerdt2015,Allerdt2016}, 
to study the many-body ground state of a quantum ($S=1/2$) impurity (modeled as an Anderson impurity) coupled to 
the edge of a zigzag nanoribbon of stanene, a slightly buckled (non-planar) honeycomb lattice of Sn atoms, 
which hosts a topologically protected metallic edge state. 
The main motivation was to study the detailed spatial structure of the spin correlations between the 
quantum impurity and the electrons in the host, which characterize the Kondo ground state. We identified 
marked differences between the results for the two distinct sites in the zigzag edge, namely 
an outermost and an innermost one, 
which we dubbed crest and trough sites, respectively. 

The behavior observed through the spin correlations is quite complex and rich. 
For substitutional impurities located at either crest or trough sites, the spin 
correlates primarily with electrons along the edge, and decouples from the bulk. 
Furthermore, irrespective of the position 
of the impurity (crest or trough site), spin correlations with conduction electrons located at crest 
sites are larger than for trough sites. The sign of the spin correlations is determined by the spins belonging 
to the same or opposite sublattice. In addition, for impurities at crest sites, ferromagnetism dominates, 
while the opposite occurs for trough sites.

The effects of SOI in the TI phase are mostly present for impurities sitting at crest sites, 
increasing the range of the correlations and introducing helical order along the edge that 
originates from an effective Dzyaloshinskii-Moriya interaction. 
Remarkably, the SOI does not affect the spin symmetry for an impurity on a 
trough site, another indication that edge states reside mostly on crest sites. 

Unlike previous calculations that consider the coupling of the impurity to one-dimensional effective 
modes \cite{Zitko2010,Vernek2016}, in our formulation the helical liquid arises naturally as an edge effect 
of the 2d bulk. It has been observed that helical liquids with an odd number of modes cannot be obtained 
from one-dimensional lattice models\cite{Wu2006}. In our treatment, the edge has unequivocally a single 
mode contribution, even away from the particle-hole symmetric point.

From the real space picture obtained from our method we are able to resolve 
the structure of the correlations at different sites along the zigzag edge and into the bulk. 
We find that substitutional impurities sitting at a trough-site form a localized bound state 
with conduction electrons in a small metallic puddle that leaks out of the edge.
On the other hand, a crest-site impurity scatters 
the edge state around it, resulting in the formation of a long range screening cloud along the edge.

Finally, we have used a variational estimate of the correlation energy to obtain a measure of $T_K$ in 
our finite system as a function of both $U$ and $\lambda_{\text{SO}}$. Our results show that 
the Kondo temperature for a substitutional impurity at a crest site is very weakly enhanced by the introduction 
of SOI, in agreement with numerical renormalizaton group calculations of a similar problem\cite{Zitko2011,Mastrogiuseppe2014}.

\begin{acknowledgments} 
The authors are grateful to T.~Hughes,  E.~Rossi, E.~Vernek, and R.~{\v{Z}}itko for 
	fruitful discussions on the subject matter of this work. 
AEF and AA acknowledge U.S. Department of Energy, Office of Basic Energy Sciences, for support under 
	grant DE-SC0014407 and GBM acknowledges 
the Brazilian Government for financial support through a Pesquisador Visitante Especial 
grant from the Ci\^encias Sem Fronteiras Program, from the Minist\'erio da Ci\^encia, 
Tecnologia e Inova\c{c}\~ao. 
\end{acknowledgments}

\bibliography{graphene_edges,Kondo-TI,chern,reviews,silicene,silicene-paper,notes,refs-andrew}

\begin{thebibliography}{60}
\bibitem [{not({\natexlab{a}})}]{note1}%
  \BibitemOpen
  \bibinfo {note} {On the one hand, the
  spin flipping mechanism, which is at the heart of the Kondo effect, although
  not breaking time reversal symmetry, will lead to backscattering of the
  surface/edge states. On the other hand, below the Kondo temperature, one
  expects that the screening of the magnetic impurities may effectively
  suppress the backscattering, restoring the helical state.}\BibitemShut
  {Stop}%
\bibitem [{not({\natexlab{b}})}]{note2}%
  \BibitemOpen
  \bibinfo {note} {The reason there are
  several peaks in Fig.~\ref{figure4} that are not visible in
  Figs.~\ref{figure3}(a) and \ref{figure5} is that for the LDOS calculations in
  Fig.~\ref{figure4} a much smaller imaginary part was used in the Green's
  function, which leads to a much higher energy resolution.}\BibitemShut
  {Stop}%
\bibitem [{not({\natexlab{c}})}]{note3}%
  \BibitemOpen
  \bibinfo {note} {A comprehensive study of
  this problem will be presented elsewhere and is beyond the scope of the
  present work.}\BibitemShut {Stop}%
\end{thebibliography}

\end{document}